\documentclass[a4paper]{article}

\usepackage{pxfonts}
\usepackage{bussproofs}
\usepackage{amssymb}
\usepackage{latexsym}
\usepackage{natbib}
\usepackage{url}

\usepackage{microtype}%Improves spacing between letters and words, must be loaded after fonts

\begin{document}

\title{Proof-Theoretic Semantics, a Problem with Negation and Prospects for Modality}
\author{Nils K\"urbis}
\date{}

\maketitle

\begin{center}
Published in \emph{The Journal of Philosophical Logic} 44/6 (2015): 713-727 \\ 
\url{http://dx.doi.org/10.1007/s10992-013-9310-6}\bigskip
\end{center}

\begin{abstract}
\noindent This paper discusses proof-theoretic semantics, the project of specifying the meanings of the logical constants in terms of rules of inference governing them. I concentrate on Michael Dummett's and Dag Prawitz' philosophical motivations and give precise characterisations of the crucial notions of harmony and stability, placed in the context of proving normalisation results in systems of natural deduction. I point out a problem for defining the meaning of negation in this framework and prospects for an account of the meanings of modal operators in terms of rules of inference.\bigskip
\end{abstract}

\begin{flushright}
\emph{et quod vides perisse perditum ducas}

Catullus
\end{flushright}

\section{Definitions and Rules of Inference}
Frege commented on definitions by abstraction in typically understated fashion: it `may be an unusual kind of definition, which presumably hasn't yet received sufficient attention from logicians; but some examples should show that it is not outrageous' (\cite{fregegrundlagen}: \S 63). Something similar can be said about definitions by rules of inference. They, too, are under-appreciated and deserve more recognition amongst a wider audience of philosophers. The logical constants, expressions like `and', `or', `not', `if-then', demonstrate best how such definitions work, their limits and avenues of further development. The approach to defining the meanings of logical constants by rules of inference stems from Gentzen's work on natural deduction. In such a calculus, each logical constant is governed by introduction rules that specify under which conditions a formula with that constant as the main operator may be derived, and elimination rules that specify what may be derived from such a formula. Gentzen's idea is embryonic, but clear enough. He noticed a `remarkable systematic' in the `inference patterns' for the logical constants and suggested that `by making these thoughts more precise it should be possible to establish on the basis of certain requirements that the elimination rules are functions of the corresponding introduction rules.' (\cite{gentzenuntersuchungen}: 189) This led him to put forward what might be called `Gentzen's Thesis': `The introductions constitute, so to speak, the ``definitions'' of the symbols concerned, and the eliminations are in the end only consequences thereof, which could be expressed thus: In the elimination of a symbol, the formula in question, whose outer symbol it concerns, may only ``be used as that which it means on the basis of the introduction of this symbol''.' (\emph{ibid.}) Gentzen's idea inspired the development of \emph{proof-theoretic semantics}. In the following, I'll largely restrict myself to Michael Dummett's and Dag Prawitz' approach, as it is the philosophically most comprehensive one.  

Prawitz' \emph{Natural Deduction} provided the formal results proof-theoretic semantics builds on. Most philosophical issues are discussed in Dummett's \emph{The Logical Basis of Metaphysics}. Dummett and Prawitz disagree occasionally. Prawitz has developed his slightly diverging views in numerous papers. But they agree on almost everything, as Dummett himself says, and as becomes clear in Prawitz' review of \emph{The Logical Basis of Metaphysics}, his contributions to various \emph{Festschriften} for Dummett and Dummett's responses.\footnote{Here is a representative selection of Prawitz' papers: (\cite{prawitzgeneral}), (\cite{prawitzmeaningandcompleteness}), (\cite{prawitzdummett}), (\cite{prawitzdummett2}), (\cite{prawitzreviewLBM}), (\cite{prawitzmeaningviaproofs}),  (\cite{prawitzdummett3}).}

I need to make a few remarks on the larger context of Dummett's and Prawitz' project. I'll keep the presentation as independent as possible, but it is occasionally important to keep in mind that some aspects of the account need to be seen in the light of that context, and these may remain less perspicuous than the broad outline I can provide here. According to Dummett and Prawitz, proof-theoretic semantics comes with another project: the justification of deduction. The aim is to impose restrictions on legitimate rules of inference, so as to narrow down the admissible rules and to single out one logic as the correct one. This connects to a further project: the correct logic gives us clues about the nature of reality. It is the scaffolding of reality into which everything else fits. Thus, proof-theoretic semantics forms part of a logical basis of metaphysics. Obviously, this leaves some assumptions implicit. The literature referred to here contains Dummett's and Prawitz' discussion of their general outlook, as well as discussions by their followers about how central these assumptions are to making the project interesting. Be that as it may, Dummett and Prawitz have given us a remarkable and grand project. It is justified to make this paper as much about their specific view as it is about a general philosophical position that is attractive to philosophers who may not share all their assumptions. 

As I'll focus on Dummett and Prawitz, I'll only discuss Gentzen's calculi of natural deduction, their preferred formal framework. Thus, I'll leave out sequent calculi with multiple conclusions and other calculi that have been developed. Dummett argues that multiple conclusions presuppose an understanding of disjunction, which is supposed to be given by rules of inference (\cite{dummettLBM}: 186f), and it would take us too far to assess to what extend other calculi can be said to satisfy Dummett's and Prawitz' restrictions, which are motivated by rather intricate philosophical considerations. I also won't talk about structural rules for the manipulation of assumptions or discharge functions, as Dummett and Prawitz never discuss these either.\footnote{This is an interesting omission: in the systems of (\cite{gentzenwiderspruchsfreiheit}), structural rules are as natural a part of natural deduction as inference rules.}

\section{Grounds and Consequences}
According to Dummett, two features central to the use of expressions determine their meanings. `The first category [of principles governing our linguistic practice] consists of those that have to do with the circumstances that warrant an assertion [$\ldots$ :] we need to know when we are \emph{entitled} to make any given assertion, and when we are \emph{required} to acknowledge it as true. [$\ldots$ Furthermore,] in acquiring language, we learn a variety of principles determining the consequences of possible utterances' (\cite{dummettLBM}: 211f). This is supposed to be true for all kinds of expressions, but it is particularly clear for the logical constants: the first feature of their use corresponds to applications of $I$-rules in a calculus of natural deduction, the second one to applications of $E$-rules. 

Dummett explains \emph{harmony} informally as a relation that ought to hold between these two features of the use of expressions. `The two complementary features of any practice ought to be in harmony with each other [$\ldots$] The notion of harmony is difficult to make precise but intuitively compelling: it is obviously not possible for the two features of the use of any expression to be determined quite independently. Given what is conventionally accepted as serving to establish the truth of a given statement, the consequences cannot be fixed arbitrarily; conversely, given what accepting a statement as true is taken to involve, it cannot be arbitrarily determined what is to count as establishing it as true.' (\cite{dummettLBM}: 215) In the case of the logical constants, the grounds for asserting a formula with main operator $\delta$, i.e. the conditions under which an $I$-rule for $\delta$ can be applied, should match, in some way to be made precise, the consequences of asserting a formula with main operator $\delta$, i.e. the conditions under which an $E$-rule for $\delta$ can be applied. The converse should also hold, a condition Dummett calls \emph{stability}. Harmony obtains if the grounds for asserting a formula with $\delta$ as main operator match the consequences of accepting it, and stability obtains if the converse also holds (cf. \cite{dummettLBM}: 287f). Thus the $I$-rules determine the $E$-rules for $\delta$, and the $E$-rules determine the $I$-rules. 

Dummett proposes two ways of making the notion of harmony precise (\cite{dummettLBM}: 250). One is in terms of the forms of rules of inference and builds on the results of (\cite{prawitznaturaldeduction}) on the normalisation of deductions. This is the topic of the next sections. The other way is in terms of \emph{conservative extensions}: `We saw that harmony, in the general sense, obtains between the verification-conditions or application-conditions of a given expression and the consequences of applying it when we cannot, by appealing to its conventionally accepted application conditions and invoking the conventional consequences of applying it, establish as true some statement which we should have had no other means of establishing: in other words, when the language is, in a transferred sense, a conservative extension of what remains of it when the given expression is deleted from its vocabulary.' (\cite{dummettLBM}: 247) I won't say much more about conservativeness, except that Dummett conjectures that, once formal conditions are in place, harmony entails conservativeness (\cite{dummettLBM}: 290).\footnote{It is sometimes said that there are inconsistent, i.e. non-conservative, connectives that have stable rules. This is impossible on the account of next section: in a logic with only stable rules, neither $\bot$ nor an arbitrary atomic proposition is provable. A version of Dummett's conjecture also holds, but it has to be said that it is not the most interesting result.}

According to Dummett, all linguistic practices ought to be harmonious. One might think that stability gives a criterion for distinguishing logical constants from other expressions: a logical constant is the kind of expression for which our practice is stable. Dummett's thought, however, would be that logical constants are expressions where \emph{all} that is needed to determine their meanings are stable rules: a logical constant is an expression the rules of which do not refer to anything else.\footnote{There are similarities to the view that logic is `topic neutral' or `carries no information'.} The meaning of a word like `red' cannot be specified by rules alone. It requires in addition reference to a practice involving red things and other colours. For Dummett, an expression that is not conservative is objectionable. `A simple case would be that of a pejorative term, e.g. ``Boche''. The condition for applying the term to someone is that he is of German nationality; the consequences of its application are that he is barbarous and more prone to cruelty than other Europeans. [...] The addition of the term ``Boche'' to a language which did not previously contain it would produce a non-conservative extension.' (\cite{dummettfregelanguage}: 454) A lack of stability is also a defect, but presumably not one that is quite so detrimental as lack of conservativeness. 

It follows that there are restrictions on admissible rules of inference. They exclude Arthur Prior's counterexample to a na\"ive view on how meaning is conferred onto logical constants by rules of inference:\footnote{(\cite{priorrunabout}). (\cite{belnaptonk}) suggests that logical constants need to be conservative, but Prior was not convinced this is sufficient (\cite{priorcontonktion}).} 

\begin{center}
$tonk I$: \AxiomC{$A$}
\UnaryInfC{$A tonk B$}
\DisplayProof\qquad\qquad $tonk E$: 
\AxiomC{$A tonk B$}
\UnaryInfC{$B$}
\DisplayProof
\end{center} 

\noindent If \emph{tonk} really was a logical constant, this would have disastrous consequences: everything would follow from everything, and although it would remove once and for all any \emph{falsche Spitzfindigkeit} from logic, it is undesirable for independent reasons. \emph{tonk} does not satisfy the criterion that the grounds of asserting $A tonk B$ match the consequences of asserting it. 

It is sometimes said that \emph{tonk} is a perfectly good logical constant and with a perfectly good meaning that is defined by its rules of inference and that the problem with it just that you wouldn't want it in your logic. It is fair to say that this is not a view I've seen in print. But it is implausible anyway. If the meaning of a logical constant is given by rules of inference only if they are in harmony, then the meaning of \emph{tonk} is not given by its rules of inference. If a logical constant is an expression such that its meaning is given purely by stable rules of inference, \emph{tonk} is not a logical constant. 

To make these thoughts a little more precise and exemplify how they apply to the logical constants, let's consider two examples. First, conjunction:   

\begin{center}
$\& I$: \AxiomC{$A$} 
\AxiomC{$B$}
\BinaryInfC{$A\& B$}
\DisplayProof\qquad\qquad $\& E$: 
\AxiomC{$A\& B$}
\UnaryInfC{$A$}
\DisplayProof\qquad 
\AxiomC{$A\& B$} 
\UnaryInfC{$B$}
\DisplayProof
\end{center}
 
\noindent The $I$-rule specifies under which conditions $A\& B$ follows, and the $E$-rules specify what follows from $A\& B$. The rules exhibit a nice balance: the $E$-rules allow us to retrieve from $A\& B$ exactly what is needed to derive $A\& B$, namely $A$ and $B$. If $\& E$ is applied directly after $\& I$, we get back to where we started. We can rearrange the deduction on the left to form the simpler one on the right:

\begin{center}
\AxiomC{$\Pi$}
\noLine
\UnaryInfC{$A$}
\AxiomC{$\Sigma$}
\noLine
\UnaryInfC{$B$}
\BinaryInfC{$A\&B$}
\UnaryInfC{$A$}
\noLine
\UnaryInfC{$\Xi$}
\DisplayProof \qquad $\leadsto$ \qquad\AxiomC{$\Pi$}
\noLine
\UnaryInfC{$A$}
\noLine
\UnaryInfC{$\Xi$}
\DisplayProof
\end{center}

\noindent Applying $\& I$ followed by $\& E$ produces an unnecessarily convoluted proof. 

As a second example, take disjunction: 

\begin{center}
$\lor I$: 
\AxiomC{$A$}
\UnaryInfC{$A\lor B$}
\DisplayProof\qquad
\AxiomC{$B$}
\UnaryInfC{$A\lor B$}
\DisplayProof\qquad\qquad $\lor E$: 
\AxiomC{$A\lor B$}
\AxiomC{}
\RightLabel{$_i$}
\UnaryInfC{$A$}
\noLine
\UnaryInfC{$\Pi$}
\noLine
\UnaryInfC{$C$}
\AxiomC{}
\RightLabel{$_i$}
\UnaryInfC{$B$}
\noLine
\UnaryInfC{$\Sigma$}
\noLine
\UnaryInfC{$C$}
\RightLabel{$_i$}
\TrinaryInfC{$C$}
\DisplayProof
\end{center}

\noindent This is a little more complicated, but in this case, too, we get back to where we started, if $\lor I$ is applied directly before $\lor E$. To apply the $E$-rule, you need a sub-deduction of a formula $C$ from $A$ and one from $B$. If the $E$-rule is applied directly after an $I$-rule, you'll get one such deduction back. We can rearrange the deduction on the left to form the one on the right: 

\begin{center}
\AxiomC{$\Sigma$}
\noLine
\UnaryInfC{$B$}
\UnaryInfC{$A\lor B$}
\AxiomC{}
\RightLabel{$_i$}
\UnaryInfC{$A$}
\noLine
\UnaryInfC{$\Pi_1$}
\noLine
\UnaryInfC{$C$}
\AxiomC{}
\RightLabel{$_i$}
\UnaryInfC{$B$}
\noLine
\UnaryInfC{$\Pi_2$}
\noLine
\UnaryInfC{$C$}
\RightLabel{$_i$}
\TrinaryInfC{$C$}
\noLine
\UnaryInfC{$\Pi$}
\DisplayProof \qquad $\leadsto$ \qquad
\AxiomC{$\Sigma$}
\noLine
\UnaryInfC{$B$}
\noLine
\UnaryInfC{$\Pi_2$}
\noLine
\UnaryInfC{$C$}
\noLine
\UnaryInfC{$\Xi$}
\DisplayProof
\end{center}

\noindent Applying $\lor E$ directly after applying $\lor I$ is an unnecessary complication. 

Following Prawitz, we call a formula that is the conclusion of an $I$-rule and major premise of an $E$-rule for its main operator a \emph{maximal formula}. Following Dummett, we call the context in which such a formula occurs a \emph{local peak}. As $\lor E$ requires the deduction of a formula $C$ as minor premise, this creates further possibilities for unnecessarily convoluted deductions. The conclusion of the rule is part of a sequence of formulas of the same shape. If the last formula is the major premise of an $E$-rule and the first one the conclusion of an $I$-rule for its main operator, this, too, should be a needless complication. We may call this a \emph{maximal sequence}. Maybe \emph{local ridge} is an adequate term for its context. It can be removed by pushing the application of the $E$-rule up until it immediately follows the application of the $I$-rule, and the maximal sequence is reduced to a maximal formula. 

On the basis of \emph{reduction procedures} for levelling local peaks and ridges, (\cite{prawitznaturaldeduction}) establishes that deductions in various logics can be \emph{normalised} and put into \emph{normal form}, so that any maximal formulas and sequences have been removed from them. A proof in normal form is particularly direct: \emph{er macht keine Umwege}, as Gentzen put it---it makes no detours.

\section{Stability, Harmony, Normalisation}
\subsection{Stability}
I'll now give a precise characterisation of stability that allows us to determine $I$-rules from $E$-rules and conversely. The rules for $\lor$ and $\&$ are quite different, so there are two different kinds of rules. In each case, I'll give only one direction of determining rules and leave the other to the reader. I define that a connective is governed by \emph{stable} rules of inference if and only if they are of type 1 or 2, with $E$/$I$-rules determined from an $I$/$E$ rule in the way to be specified, and there are no conditions on the application of the rules. It would be natural to stipulate that in rules of type 1, the $I$-rules specify the meanings of the logical constants, whereas it is the $E$-rules in type 2. It should be kept in mind, though, that it is not just the $E$-rules or just the $I$-rules that specify meanings. Rather, it is the $I$/$E$-rules plus a principle of stability, on the basis of which the correct $E$/$I$-rules are determined. In a sense, then, it does not matter which rules we pick as the ones determining meaning. 

We stipulate that for a constant to be governed by rules of type 1 it has exactly one $I$-rule, which can be of any form whatsoever, as long as the conclusion of the rule is constructed by connecting all and only the premises and discharged hypotheses by using the logical constant as main operator. 

We `read off' the $E$-rules for the constant from its $I$-rule in this way: To each premise of the $I$-rule there corresponds an $E$-rule which has that premise of the $I$-rule as its conclusion and if there are discharged hypotheses above that premise of the $I$-rule, these become minor premises of the $E$-rule. The major premise of the $E$-rule is of course the conclusion of the $I$-rule. 

Here are some examples. $\& I$ has two premises and no discharged hypotheses above either of them, so it has two $E$-rules, each without minor premises. $\supset$ has an $I$-rule with one premise and one discharged hypothesis above it, so it has one $E$-rule with one minor premise:

\begin{center}
$\supset I$: \AxiomC{}
\RightLabel{$_i$}
\UnaryInfC{$A$}
\noLine
\UnaryInfC{$\Pi$}
\noLine
\UnaryInfC{$B$}
\RightLabel{$_i$}
\UnaryInfC{$A\supset B$}
\DisplayProof
\qquad\qquad $\supset E$: \AxiomC{$A\supset B$}
\AxiomC{$A$}
\BinaryInfC{$B$}
\DisplayProof
\end{center}

\noindent \textbf{T} has an $I$-rule with one premise and no discharged hypotheses, so it has one $E$-rule with no minor premises: 

\begin{center}
$\textbf{T} I$: \AxiomC{$A$}
\UnaryInfC{\textbf{T}$A$}
\DisplayProof
\qquad\qquad $\textbf{T} E$: \AxiomC{\textbf{T}$A$}
\UnaryInfC{$A$}
\DisplayProof
\end{center}

\noindent A somewhat peculiar case is \emph{verum} which has the $I$-rule: 
\begin{center}
$\top I$: 
\AxiomC{}
\UnaryInfC{$\top$}
\DisplayProof
\end{center}

\noindent The rule has no premises, hence $\top$ has no $E$-rule.  

We stipulate that for a constant to be governed by rules of type 2, it has exactly one $E$-rule, which can be of any form whatsoever as long as its major premise is constructed with the logical constant as main operator from all and only the discharged hypotheses of collateral deductions of minor premises $C$, which is also the conclusion of the rule. 

We `read off' the $I$-rules for the constant from its $E$-rule in this way: To each collateral deduction of the $E$-rule there corresponds an $I$-rule which has as its premises the discharged hypotheses of that collateral deduction. The conclusion of the $I$-rule is of course the major premise of the $E$-rule. 

Here are some examples. $\lor E$ has two collateral deductions with one discharged hypothesis each, so it has two $I$-rules, each with one premise. $\times$ has an $E$-rule with one collateral deduction with two discharged hypotheses, so it has one $I$-rule with two premises:

\begin{center}
$\times E$: \AxiomC{$A\times B$}
\AxiomC{$\underbrace{\overline{\ A \ }^i\quad \overline{\ B\ }^i}$}
\noLine
\UnaryInfC{$\Pi$}
\noLine
\UnaryInfC{$C$}
\RightLabel{$_i$}
\BinaryInfC{$C$}
\DisplayProof\qquad\qquad $\times I$: 
\AxiomC{$A$}
\AxiomC{$B$}
\BinaryInfC{$A\times B$}
\DisplayProof
\end{center}

\noindent $\times I$ has the same form as $\& I$. In classical and intuitionist logic, $A\&B$ is equivalent to $A\times B$. The example shows that it is important to know which kind of rules a connective has. 

\textbf{t} has an $E$-rule with one collateral deduction with one discharged hypothesis, so it has one $I$-rule with one premise: 

\begin{center}
$\textbf{t} E$: 
\AxiomC{\textbf{t}$A$}
\AxiomC{}
\RightLabel{$_i$}
\UnaryInfC{$A$}
\noLine
\UnaryInfC{$\Pi$}
\noLine
\UnaryInfC{$C$}
\RightLabel{$_i$}
\BinaryInfC{$C$}
\DisplayProof\qquad\qquad $\textbf{t} I$: 
\AxiomC{$A$}
\UnaryInfC{\textbf{t}$A$}
\DisplayProof
\end{center}

\noindent As $\textbf{t} I$ has the same form as $\textbf{T} I$, this is another example that shows that it is important to know which kind of rules a connective has. 

Another peculiar case is $\bot$, which has an $E$-rule with no collateral deductions, so no $I$-rule: 

\begin{center}
$\bot E$: 
\AxiomC{$\bot$}
\UnaryInfC{$C$}
\DisplayProof
\end{center}

\noindent I should add that some accounts of harmony use only one kind of rules, by allowing minor premises that are not conclusions of sub-deductions into rules of type 2.\footnote{See (\cite{nissimnote}) and (\cite{readGEharmony}) and literature referred to there.}

\subsection{Harmony}
I define that a connective is governed by \emph{harmonious} rules if and only if they are of type 1 or 2, with $E$/$I$-rules determined from an $I$/$E$ rule in the way specified, and there are conditions on the application of the rules. 

As an example, the rules for necessity are of type 1 with conditions for their application, so they are harmonious but not stable:

\noindent \begin{center}
$\Box I$: \AxiomC{$B$}
\UnaryInfC{$\Box B$}
\DisplayProof\qquad\qquad $\Box E$: 
\AxiomC{$\Box A$}
\UnaryInfC{$A$}
\DisplayProof
\end{center}

\noindent For \textbf{S4}-necessity, all the formulas $B$ depends on must be of the form $\Box C$.\footnote{In an intuitionist modal logic where both $\Box$ and $\Diamond$ are present, the restrictions may have to be more complicated. For the classical case, we can assume that the logic only has $\Box$.} For \textbf{S5}-necessity, they must be \emph{modalised}, i.e. every propositional variable is in the scope of a modal operator (so $\bot$ and $\top$ are modalised). 

According to some accounts, a connective with the same $I$-rule as \& but lacking one if its $E$-rules is harmonious but not stable. Such a connective doesn't appear to me to be very interesting, and such an account of harmony strikes me as less fruitful than mine, where this connective is not harmonious. Connectives that are harmonious but not stable according to my definition are interesting and illustrate a philosophically significant feature. According to proof-theoretic semantics, if the rules for a connective are not stable, they do not determine its meaning completely. Quite plausibly that is the case for $\Box$. I cannot learn the meaning of $\Box$ by adding its rules to my repertoire if that doesn't already contain $\Box$, as I can in the case of $\&$ and $\lor$. To know under which conditions I can apply the $I$-rule, I need to have used formulas of the form $\Box C$ already, so the rule presupposes some understanding of $\Box$, wherever that understanding might come from. 

Possibility is a similar case. For \textbf{S4}-possibility, $C$ must be of form $\Diamond D$, and all formulas it depends on, except possibly $B$, must be of form $\Box E$;\footnote{I'm using this restriction as it mirrors the rules of (\cite{biermannpaiva}) in the next section. An intuitionist modal logic may require more complicated restrictions, such as those of (\cite{prawitznaturaldeduction}).} for \textbf{S5}-possibility, they are required to be modalised: 

\noindent\begin{center} 
$\Diamond E$: 
\AxiomC{$\Diamond B$}
\AxiomC{}
\RightLabel{$_i$}
\UnaryInfC{$B$}
\noLine
\UnaryInfC{$\Pi$}
\noLine
\UnaryInfC{$C$}
\RightLabel{$_i$}
\BinaryInfC{$C$}
\DisplayProof\qquad\qquad $\Diamond I$: 
\AxiomC{$A$}
\UnaryInfC{$\Diamond A$}
\DisplayProof 
\end{center}

\noindent These are rules of type 2 with conditions on their application. They are harmonious, but not stable. It is plausible that they cannot determine the meaning of $\Diamond$ completely, as $\Diamond$ is referred to in the restrictions.

\subsection{Normalisation}
Harmony and stability could be said to be properties of rules of inference. But it's a little pointless to consider rules of inference independently of a formal system of logic. The question whether a connective satisfies the criteria of proof-theoretic semantics can only be answered by considering a logic it occurs in and whether deductions in that logic normalise. It is, however, possible to give generalised reduction procedures for the removal of maximal formulas and sequences that hold for all stable rules. Thus, if all the connectives of a logic are governed by stable rules, then all deductions can be normalised. If some of the rules are merely harmonious, this question can only be decided on a case by case basis. For example, in intuitionist logic, the maximal formula $A\supset B$ in the deduction on the left can be removed by rearranging it to form the deduction on the right: 

\begin{center}
\AxiomC{}
\RightLabel{$_i$}
\UnaryInfC{$A$}
\noLine
\UnaryInfC{$\Pi$}
\noLine
\UnaryInfC{$B$}
\RightLabel{$_i$}
\UnaryInfC{$A\supset B$}
\AxiomC{$\Sigma$}
\noLine
\UnaryInfC{$A$}
\BinaryInfC{$B$}
\noLine
\UnaryInfC{$\Xi$}
\DisplayProof $\qquad$ $\leadsto$ \qquad \AxiomC{$\Sigma$}
\noLine
\UnaryInfC{$A$}
\noLine
\UnaryInfC{$\Pi$}
\noLine
\UnaryInfC{$B$}
\noLine
\UnaryInfC{$\Xi$}
\DisplayProof
\end{center}

\noindent If we add $\Box$ to intuitionist logic, the reduction procedure cannot be applied: if there are applications of $\Box I$ in $\Pi$ in the original deduction, nothing guarantees that they remain correct in the rearranged deduction. In classical logic without $\supset$ as a primitive, a similar problem arises from the rules for negation. 

To make normalisation in modal logic possible, Prawitz modifies the restriction on the application of $\Box I$ and $\Diamond E$ (\cite{prawitznaturaldeduction}: 74ff). (\cite{biermannpaiva}) formulate rules that incorporate the restrictions in the shape of $\Box I$ and $\Diamond E$: 

\begin{center} 
\def\defaultHypSeparation{\hskip .1in}
$\Box I$: \AxiomC{$A_1\ldots A_n$}
\AxiomC{$\underbrace{\overline{\ A_1 \ }^i \ldots \overline{ \ A_n \ }^i}$}
\noLine
\UnaryInfC{$\Xi$}
\noLine
\UnaryInfC{$B$}
\RightLabel{$_i$}
\BinaryInfC{$\Box B$}
\DisplayProof\qquad $\Diamond E$: 
\AxiomC{$\Diamond B$}
\AxiomC{$A_1\ldots A_n$}
\AxiomC{$\underbrace{\overline{\ B \ }^i, \overline{\ A_1 \ }^i \ldots \overline{ \ A_n \ }^i}$}
\noLine
\UnaryInfC{$\Sigma$}
\noLine
\UnaryInfC{$C$}
\RightLabel{$_i$}
\TrinaryInfC{$C$}
\DisplayProof
\end{center}

\noindent $A_1\ldots A_n$ are exactly the undischarged assumptions in $\Xi$ and $A_1\ldots A_n, B$ in $\Sigma$. For \textbf{S4}, $A_1\ldots A_n$ are required to be of the form $\Box D_1\dots \Box D_n$ and $C$ to be of the form $\Diamond D$. Biermann and de Paiva show that deductions in a system of intuitionist \textbf{S4} with these rules normalise. It can also be shown that deductions in a system of classical \textbf{S4} with the negation rules of the next section normalise. The same is true for classical and intuitionist \textbf{S5}, where the restrictions require only that $A_1\ldots A_n, B, C$ are modalised.  

These rules for $\Box$ and $\Diamond$ are neither of type 1 not of type 2. However, they merely transform rules of these types with constraints on their application into rules which show the constraints explicitly in their form. Thus I extend the definition of harmony to cover such rules, too. So although the meanings of $\Box$ and $\Diamond$ are not determined completely by the rules governing them, and an understanding of modal notions must be built on something else, they are not objectionable, as we can still give harmonious rules for them that allow the normalisation of deductions.

\section{A Problem with Negation}
We can define $\neg A$ as $A \supset \bot$. This suffices for intuitionist logic. To formalise classical logic, Prawitz adds \emph{consequentia mirabilis} to a system with $\bot$, $\&$ and $\supset$. $\supset$ can be dispensed with if we add $I$- and $E$-rules for a primitive $\neg$:  

\begin{center}
$\neg I$: 
\AxiomC{}
\RightLabel{$_i$}
\UnaryInfC{$A$}
\noLine
\UnaryInfC{$\Xi$}
\noLine
\UnaryInfC{$\bot$}
\RightLabel{$_i$}
\UnaryInfC{$\neg A$}
\DisplayProof \qquad
$\neg E$: 
\AxiomC{$\neg A$}
\AxiomC{$A$}
\BinaryInfC{$\bot$}
\DisplayProof\qquad
$cm$: 
\AxiomC{}
\UnaryInfC{$\neg A$} 
\noLine
\UnaryInfC{$\Sigma$}
\noLine
\UnaryInfC{$\bot$}
\RightLabel{$_i$}
\UnaryInfC{$A$}
\DisplayProof
\end{center}

\noindent Deductions in this system normalise. Nonetheless, the rules for classical negation are not stable, which is why Dummett and Prawitz consider them to be defective, so that the meaning of classical negation cannot be given by rules governing it. \emph{Consequentia mirabilis} introduces grounds for asserting propositions that are not justified relative to their consequences. They would not be there in the absence of that rule and are not matched by the propositions' consequences. \emph{Consequentia mirabilis} allows us to assert propositions more often than we should be allowed, given their consequences. 

This disharmony in the rules of classical negation has almost disastrous consequences. Dummett's and Prawitz' argument to that effect is rather intricate and built on the idea that the meaning of expressions is tied to their use and that a theory of meaning is a theory of understanding, but maybe the following will do for my purposes.\footnote{Besides the works already cited, (\cite{dummettmeaningI}), (\cite{dummettmeaningII}),  (\cite{dummettjustdeduction}) and (\cite{dummettphilbasisint}) contain extensive discussions of the issue.} The meaning of $\neg\neg A$ is dependent on the meaning of $A$ and $\neg$. It may happen that a sentence of a language where double negation elimination is employed can be verified only via its double negation. In such a case the move from $\neg\neg A$ to $A$ would contribute to the meaning of $A$, because it licenses assertions of $A$ not otherwise possible, as \emph{ex hypothesi} no other verification is available. Hence the meaning of $A$ would depend on the meaning of $\neg\neg A$. This is a circular dependence of meaning and hence $A$ cannot have a coherent meaning at all. A speaker could not break into the circle and learn the meaning of $A$. Thus, using classical negation, a large range of propositions literally become meaningless.

There are, of course, all kinds of ways in which one can get classical logic from intuitionist logic by adding axioms or rules of inference. According to Dummett and Prawitz, all these rules and axioms are defective. In fact, it is not the choice of rules and axioms that matters. The problem is $\neg\neg A\vdash A$, no matter how it is derived.\footnote{There are several proposals for how to formulate classical logic in a such a way that its rules may count as harmonious, e.g. (\cite{milneharmony}) and (\cite{readharmony}). It is fair to say that they all deviate in some way from Dummett's and Prawitz' harmony.} More interesting than considering alternative ways of formalising classical logic is the question whether the meaning of intuitionist negation can reasonably be said to be determined by rules of inference. Because if it cannot, the objection that the meaning of classical negation cannot be determined by rules of inference rather loses its force. 

The sole rule governing $\bot$ allows us to infer anything whatsoever from it, where we can restrict the conclusion to atomic propositions.\footnote{Or $\Box A$, in modal logic. A local peak with $\Box E$ after $\bot E$ can be levelled.} It is supposed to confer on $\bot$ the meaning of a proposition that is always false. But assume all atomic propositions are true. Then $\bot$ doesn't have to be false. It may be true. So $\bot E$ does not determine the intended meaning of $\bot$. Hence the rules do not determine the intended meaning of $\neg$ either.\footnote{This argument can be found in (\cite{handnegation}), but it has probably occurred to many philosophers independently, amongst them the present author.}

This is a problem for proof-theoretic semantics and the justification of deduction. Dummett and Prawitz object to classical negation that its meaning cannot be determined by rules of inference. In fact intuitionist negation is in the same boat. 

As Dummett thinks that our practice may be mistaken and stand in need of revision, one response is to revise the intuition that $A$ and $\neg A$ cannot be both true: there is one odd case in which they are. It is difficult to say whether this is satisfactory. Revisionism comes to an end somewhere. There are methodological issues the response would need to address, but as it doesn't seem to be very popular, let's leave it at that. 

A more prominent response is to introduce a primitive notion of \emph{incompatibility} and define negation in terms of it. Neil Tennant has proposed that this relation holds between facts such as $a$'s being red and $a$'s being green, $a$'s being here and $a$'s being over there, $a$'s once being the case and $a$'s only going to be the case. Then $\neg A$ is true if $A$ entails some sentences that assert incompatible facts.\footnote{(\cite{tennantnegation}), which is a response to (\cite{handnegation}).} Robert Brandom uses similar examples, but characterises incompatibility as a relation between sentences.\footnote{(\cite{brandomlocke}: Lecture 5, p.8ff). His definition of negation is different from Tennant's, but there is no need to go into the details here. A similar approach can already be found in (\cite{demos}). It never really caught on, maybe because of Russell's criticism (\cite{russellpropositions}: 5ff), (\cite{russelllogicalatomism}: 211ff).} I don't find the approach convincing. It is not obvious to some metaphysicians whether the examples really are incompatibilities, and it doesn't strike me as desirable to make the definition of negation dependent on the outcomes of arcane debates in metaphysics. Even if we ignore these metaphysicians, there is a more significant problem. I don't know how to go on, to apply this primitive notion of incompatibility in new cases and come up with new examples. For instance, is finding beetroot delicious incompatible with being me? Is being a dinosaur incompatible with being a reptile? I have difficulties coming up with examples of primitive incompatibilities involving that bottle of Vin Jaune on my desk that aren't like the ones I've already seen in the literature using colour, place or time. I'm not convinced that's just lack of imagination. I know a lot about what this bottle is not, but I don't know a lot about what its properties are primitively incompatible with. Take shape. Brandom and Tennant hold that being square is primitively incompatible with being round. But why is that incompatibility primitive, rather than something that follows because `$a$ is round' and `$a$ is square' entails a contradiction? I understand negation a lot better than incompatibility. The latter strikes me as rather more complex than the former. In fact, I think I only understand it because I can define it in terms of negation: $p$ and $q$ are incompatible if and only if $p$ and $q$ cannot both be true.

Tennant and Brandom, incidentally, have similar inferentialist views, but the former argues that negation defined in terms of incompatibility is intuitionist, and the latter argues it is classical. This suggests that it is not clear what the meaning of a negation defined in terms of incompatibility actually is. 

Huw Price has proposed a definition of negation in terms of two fundamental speech acts, assertion and denial, that are incompatible. A proposition cannot be asserted and denied at the same time. Arguably, the resulting negation is classical.\footnote{See (\cite{pricesense}), (\cite{pricewhynot}), (\cite{pricenotagain}). (\cite{rumfittyesno}) develops a formal framework for the account, which has sparked some discussion. See (\cite{dummettyesno}), (\cite{gibbard}), (\cite{rumfittreply1}), (\cite{ferreira}), (\cite{rumfittreply2}). (\cite{textordenial}) is a critical assessment of whether there is a speech act of denial that is prior to the assertion of negated sentences. It is worth adding that according to Tennant and Rumfitt, $\bot$ isn't a proposition, but a `structural punctuation mark'. Nonetheless, it plays a role in their calculi that can be played by a proposition or speech act. Despite their declarations, someone who is being taught the rules of their calculi may not come to the conclusion that $\bot$ is quite so special, and I'm not convinced that would just be a confusion.} The approach settles the question of the justification of deduction in an interestingly different way from Dummett's and Prawitz', and so I won't go into any further details. 

A final response which I consider to be closest to Dummett's and Prawitz' position is based on a suggestion by Tennant. A language could not be learnt if all sentences in it were true and never changed truth value. For language to be possible, the contrast between sentences being true and being false is necessary. This response imports further considerations from the philosophy of language into proof-theoretic semantics, which is of course also where the requirement of harmony comes from. These may have metaphysical consequences, but the metaphysics is not the starting point. The approach, however, appeals to truth and falsity as primitives, which may make it impossible to solve the question whether classical or intuitionist logic is correct. Something needs to be said about the relation between truth and falsity. In order not to prejudge the issue, that relation would have to be thin enough so as not to decide whether every proposition is determinately either true or false. I'm not saying this can't be done, but it is certainly a challenge.

\section{Prospects for Modality}
Some notions of necessity and possibility are purely logical. Unless proof-theoretic semantics has an account of their meaning, it is seriously incomplete. In this section, I'll give the beginnings of a response to the unsatisfactory state of affairs that arises because $\Box$ and $\Diamond$ aren't governed by stable rules. It is based on the following observation. A crucial assumption of proof-theoretic semantics is that speakers can follow rules of inference. This ability imparts understanding of the meanings of the logical constants onto speakers. If we spell out in some more detail what this entails, we can see that proof-theoretic semantics assumes that speakers implicitly understand certain modal notions. This opens the door to an account of the meanings of modal operators within proof-theoretic semantics.\footnote{Modal notions are almost never discussed in Dummett/Prawitz-style proof-theoretic semantics. (\cite{readharmonymodality}) uses a labelled deductive system, which strikes me as possible worlds semantics dressed up. (\cite{pfenningdaviesmodal}) is closer to what I say here.}  

Proof-theoretic semantics assumes that speakers can draw logical inferences. Logical inferences establish a \emph{necessary connection} between premises and conclusion. Thus proof-theoretic semantics assumes that speakers grasp a relative notion of necessity, namely the notion of a necessary connection between premises and conclusions of the rules of inference governing logical constants. It is built into one of the primitives of proof-theoretic semantics. This observation motivates the introduction of a connective that captures this relative notion of necessity in the language. Its meaning would not be determined completely by rules of inference governing it, as it captures what speakers are implicitly assumed to understand already. Thus we need not require its rules to be stable, but harmonious rules would suffice. 

Proof-theoretic semantics also assumes that speakers can draw inferences from propositions assumed for the sake of the argument. Making an assumption sometimes amounts to considering a possibility. We can assume impossibilities, thus, making an assumption is a wider notion than considering a possibility. This should lend itself to an account of a notion of possibility that is dependent on relative necessity, as assuming something is dependent on drawing inferences. Thus the rules for an operator capturing this notion of possibility also need not be stable and they may refer to the relative notion of necessity. 

Proof-theoretic semantics assumes that speakers have the conceptual resources to understand certain modal notions. What is needed is a formal system that captures these implicit modal notions explicitly in modal operators. Such a system constitutes an account of the meanings of modal operators in the framework of proof-theoretic semantics. It could be attractive to a wide range of philosophers, as it promises an account of modality that avoids reference to possible worlds, which is metaphysically unattractive.\footnote{I would like to thank Anneli Jefferson, Jessica Leech and Julien Murzi for their very helpful comments on this paper.}

\bigskip 

\setlength{\bibsep}{0pt}
\bibliographystyle{chicago}
\bibliography{KurbisPTS}

\begin{thebibliography}{}

\bibitem[\protect\citeauthoryear{Belnap}{Belnap}{1962}]{belnaptonk}
Belnap, N. (1962).
\newblock Tonk, plonk and blink.
\newblock {\em Analysis\/}~{\em 22}, 130--134.

\bibitem[\protect\citeauthoryear{Biermann and de~Paiva}{Biermann and
  de~Paiva}{2000}]{biermannpaiva}
Biermann, G.~M. and V.~C.~V. de~Paiva (2000).
\newblock On an intuitionistic modal logic.
\newblock {\em Studia Logica\/}~{\em 65}, 383--416.

\bibitem[\protect\citeauthoryear{Brandom}{Brandom}{2006}]{brandomlocke}
Brandom, R. (2006).
\newblock {The John Locke Lectures}.
\newblock \url{http://www.pitt.edu/~brandom/locke/index.html} (accessed:
  08/11/2013).

\bibitem[\protect\citeauthoryear{Demos}{Demos}{1917}]{demos}
Demos, R. (1917).
\newblock A discussion of a certain type of negative propositions.
\newblock {\em Mind\/}~{\em 26}, 188--196.

\bibitem[\protect\citeauthoryear{Dummett}{Dummett}{1978a}]{dummettjustdeduction}
Dummett, M. (1978a).
\newblock The justification of deduction.
\newblock In {\em Truth and Other Enigmas}, pp.\  290--318. London: Duckworth.

\bibitem[\protect\citeauthoryear{Dummett}{Dummett}{1978b}]{dummettphilbasisint}
Dummett, M. (1978b).
\newblock The philosophical basis of intuitionistic logic.
\newblock In {\em Truth and Other Enigmas}, pp.\  215--247. London: Duckworth.

\bibitem[\protect\citeauthoryear{Dummett}{Dummett}{1981}]{dummettfregelanguage}
Dummett, M. (1981).
\newblock {\em Frege. Philosophy of Language\/} (2 ed.).
\newblock London: Duckworth.

\bibitem[\protect\citeauthoryear{Dummett}{Dummett}{1993a}]{dummettLBM}
Dummett, M. (1993a).
\newblock {\em The Logical Basis of Metaphysics}.
\newblock Cambridge, Mass.: Harvard University Press.

\bibitem[\protect\citeauthoryear{Dummett}{Dummett}{1993b}]{dummettmeaningI}
Dummett, M. (1993b).
\newblock {What is a Theory of Meaning? (I)}.
\newblock In {\em The Seas of Language}, pp.\  1--33. Oxford: Clarendon.

\bibitem[\protect\citeauthoryear{Dummett}{Dummett}{1993c}]{dummettmeaningII}
Dummett, M. (1993c).
\newblock {What is a Theory of Meaning? (II)}.
\newblock In {\em The Seas of Language}, pp.\  34--93. Oxford: Clarendon.

\bibitem[\protect\citeauthoryear{Dummett}{Dummett}{2002}]{dummettyesno}
Dummett, M. (2002).
\newblock {``Yes'', ``No'' and ``Can't Say''}.
\newblock {\em Mind\/}~{\em 111}, 289--295.

\bibitem[\protect\citeauthoryear{Ferreira}{Ferreira}{2008}]{ferreira}
Ferreira, F. (2008).
\newblock The co-ordination principles: A problem for bilateralism.
\newblock {\em Mind\/}~{\em 117}, 1051--1057.

\bibitem[\protect\citeauthoryear{Francez and Dyckhoff}{Francez and
  Dyckhoff}{2012}]{nissimnote}
Francez, N. and R.~Dyckhoff (2012).
\newblock A note on harmony.
\newblock {\em Journal of Philosophical Logic\/}~{\em 41}, 613--628.

\bibitem[\protect\citeauthoryear{Frege}{Frege}{1990}]{fregegrundlagen}
Frege, G. (1990).
\newblock {\em Die Grundlagen der Arithmetik}.
\newblock Hildesheim, Z\"urich, New York: Olms.

\bibitem[\protect\citeauthoryear{Gentzen}{Gentzen}{1934}]{gentzenuntersuchungen}
Gentzen, G. (1934).
\newblock {Untersuchungen \"uber das logische Schlie\ss en}.
\newblock {\em Mathematische Zeitschrift\/}~{\em 39}, 176--210, 405--431.

\bibitem[\protect\citeauthoryear{Gentzen}{Gentzen}{1935}]{gentzenwiderspruchsfreiheit}
Gentzen, G. (1935).
\newblock {Die Widerspruchsfreiheit der reinen Zahlentheorie}.
\newblock {\em Mathematische Annalen\/}~{\em 112}, 493--565.

\bibitem[\protect\citeauthoryear{Gibbard}{Gibbard}{2002}]{gibbard}
Gibbard, P. (2002).
\newblock Price and {Rumfitt} on rejective negation and classical logic.
\newblock {\em Mind\/}~{\em 111}, 297--303.

\bibitem[\protect\citeauthoryear{Hand}{Hand}{1999}]{handnegation}
Hand, M. (1999).
\newblock Anti-realism and falsity.
\newblock In D.~Gabbay and H.~Wansing (Eds.), {\em What is Negation?}, pp.\
  185--198. Dortrecht: Kluwer.

\bibitem[\protect\citeauthoryear{Milne}{Milne}{1994}]{milneharmony}
Milne, P. (1994).
\newblock Classical harmony: Rules of inference and the meanings of the logical
  constants.
\newblock {\em Synthese\/}~{\em 100}, 49--94.

\bibitem[\protect\citeauthoryear{Pfenning and Davies}{Pfenning and
  Davies}{2001}]{pfenningdaviesmodal}
Pfenning, F. and R.~Davies (2001).
\newblock A judgemental reconstruction of modal logic.
\newblock {\em Mathematical Structures in Computer Science\/}~{\em 11},
  511--540.

\bibitem[\protect\citeauthoryear{Prawitz}{Prawitz}{1965}]{prawitznaturaldeduction}
Prawitz, D. (1965).
\newblock {\em Natural Deduction}.
\newblock Stockholm, G\"oteborg, Uppsala: Almqvist and Wiksell.

\bibitem[\protect\citeauthoryear{Prawitz}{Prawitz}{1974}]{prawitzgeneral}
Prawitz, D. (1974).
\newblock On the idea of a general proof-theory.
\newblock {\em Synthese\/}~{\em 27}, 63--77.

\bibitem[\protect\citeauthoryear{Prawitz}{Prawitz}{1979}]{prawitzmeaningandcompleteness}
Prawitz, D. (1979).
\newblock Proofs and the meaning and completeness of the logical constants.
\newblock In J.~H. \emph{et al.} (Ed.), {\em Essays on Mathematical and
  Philosophical Logic}, pp.\  25--40. Dordrecht: Reidel.

\bibitem[\protect\citeauthoryear{Prawitz}{Prawitz}{1987}]{prawitzdummett}
Prawitz, D. (1987).
\newblock Dummett on a theory of meaning and its impact on logic.
\newblock In B.~Taylor (Ed.), {\em Michael Dummett: Contributions to
  Philosophy}, pp.\  117--165. Dordrecht: Nijhoff.

\bibitem[\protect\citeauthoryear{Prawitz}{Prawitz}{1994a}]{prawitzdummett2}
Prawitz, D. (1994a).
\newblock Meaning theory and anti-realism.
\newblock In B.~McGuiness (Ed.), {\em The Philosophy of Michael Dummett}, pp.\
  79--89. Dordrecht: Kluwer.

\bibitem[\protect\citeauthoryear{Prawitz}{Prawitz}{1994b}]{prawitzreviewLBM}
Prawitz, D. (1994b).
\newblock Review of \emph{The Logical Basis of Metaphysics}.
\newblock {\em Mind\/}~{\em 103}, 373--376.

\bibitem[\protect\citeauthoryear{Prawitz}{Prawitz}{2006}]{prawitzmeaningviaproofs}
Prawitz, D. (2006).
\newblock Meaning approached via proofs.
\newblock {\em Synthese\/}~{\em 148}, 507--524.

\bibitem[\protect\citeauthoryear{Prawitz}{Prawitz}{2007}]{prawitzdummett3}
Prawitz, D. (2007).
\newblock Pragmatist and verificationist theories of meaning.
\newblock In R.~Auxier and L.~Hahn (Eds.), {\em The Philosophy of Michael
  Dummett}, pp.\  455--481. Chicago: Open Court.

\bibitem[\protect\citeauthoryear{Price}{Price}{1983}]{pricesense}
Price, H. (1983).
\newblock {Sense, Assertion, Dummett and Denial}.
\newblock {\em Mind\/}~{\em 92}, 161--173.

\bibitem[\protect\citeauthoryear{Price}{Price}{1990}]{pricewhynot}
Price, H. (1990).
\newblock {Why `Not'?}
\newblock {\em Mind\/}~{\em 99}, 221--238.

\bibitem[\protect\citeauthoryear{Price}{Price}{2010}]{pricenotagain}
Price, H. (2010).
\newblock {`Not' Again}.
\newblock \url{http://prce.hu/w/preprints/NotAgain.pdf} (accessed 08/11/2013).

\bibitem[\protect\citeauthoryear{Prior}{Prior}{1961}]{priorrunabout}
Prior, A. (1961).
\newblock The runabout inference ticket.
\newblock {\em Analysis\/}~{\em 21}, 38--39.

\bibitem[\protect\citeauthoryear{Prior}{Prior}{1964}]{priorcontonktion}
Prior, A. (1964).
\newblock Conjunction and contonktion revisited.
\newblock {\em Analysis\/}~{\em 24}, 191--195.

\bibitem[\protect\citeauthoryear{Read}{Read}{2000}]{readharmony}
Read, S. (2000).
\newblock Harmony and autonomy in classical logic.
\newblock {\em Journal of Philosophical Logic\/}~{\em 29}, 123--154.

\bibitem[\protect\citeauthoryear{Read}{Read}{2008}]{readharmonymodality}
Read, S. (2008).
\newblock Harmony and modality.
\newblock In L.~K. C.~D{\'e}gremont and H.~R{\"u}ckert (Eds.), {\em Dialogues,
  Logics and Other Strange Things: Essays in Honour of Shahid Rahman}, pp.\
  285--303. London: College Publications.

\bibitem[\protect\citeauthoryear{Read}{Read}{2010}]{readGEharmony}
Read, S. (2010).
\newblock General-elimination harmony and the meaning of the logical constants.
\newblock {\em Journal of Philosophical Logic\/}~{\em 39}, 557--576.

\bibitem[\protect\citeauthoryear{Rumfitt}{Rumfitt}{2000}]{rumfittyesno}
Rumfitt, I. (2000).
\newblock {``Yes'' and ``No''}.
\newblock {\em Mind\/}~{\em 109}, 781--823.

\bibitem[\protect\citeauthoryear{Rumfitt}{Rumfitt}{2002}]{rumfittreply1}
Rumfitt, I. (2002).
\newblock Unilateralism disarmed: A reply to {Dummett and Gibbard}.
\newblock {\em Mind\/}~{\em 111}, 305--312.

\bibitem[\protect\citeauthoryear{Rumfitt}{Rumfitt}{2008}]{rumfittreply2}
Rumfitt, I. (2008).
\newblock Co-ordination principles: A reply.
\newblock {\em Mind\/}~{\em 117}, 1059--1063.

\bibitem[\protect\citeauthoryear{Russell}{Russell}{1919}]{russellpropositions}
Russell, B. (1919).
\newblock On propositions: What they are and how they mean.
\newblock {\em Proceedings of the Aristotelian Society, Supplementary
  Volumes\/}~{\em 2}, 1--43.

\bibitem[\protect\citeauthoryear{Russell}{Russell}{1956}]{russelllogicalatomism}
Russell, B. (1956).
\newblock The philosophy of logical atomism.
\newblock In {\em Logic and Knowledge. Essays 1901-1950}, pp.\  175--283.
  Routledge.

\bibitem[\protect\citeauthoryear{Tennant}{Tennant}{1999}]{tennantnegation}
Tennant, N. (1999).
\newblock Negation, absurdity and contrariety.
\newblock In D.~Gabbay and H.~Wansing (Eds.), {\em What is Negation?}, pp.\
  199--222. Dortrecht: Kluwer.

\bibitem[\protect\citeauthoryear{Textor}{Textor}{2011}]{textordenial}
Textor, M. (2011).
\newblock Is 'no' a force indicator? {No}!
\newblock {\em Analysis\/}~{\em 71\/}(3), 448--456.

\end{thebibliography}

\end{document}